\documentclass[aps,prmaterials,preprint,groupedaddress]{revtex4-2}

\usepackage{graphicx}
\usepackage{color}
\usepackage{siunitx}
\usepackage{stmaryrd}	
\usepackage{multirow}
\usepackage{bm}
\usepackage[colorlinks=true,linkcolor=black,citecolor=blue,urlcolor=blue,anchorcolor=blue]{hyperref}
\usepackage{amssymb,amsmath,amsfonts,latexsym,mathrsfs,amsthm}
\usepackage{natbib}
\usepackage{soul}

\begin{document}


\title{Enhanced functional reversibility in lead-free ferroelectric material over long cycle pyroelectric energy conversion}



\author{Chenbo Zhang}
\email{cbzhang@tongji.edu.cn}
\affiliation{MOE Key Laboratory of Advanced Micro-Structured Materials, School of Physics Science and Engineering, Institute for Advanced Study, Tongji University, Shanghai, 200092, China}

\author{Zeyuan Zhu}
\affiliation{Department of Mechanical and Aerospace Engineering, Hong Kong University of Science and Technology, Kowloon, Hong Kong SAR}

\author{Ka Hung Chan}
\affiliation{Department of Mechanical and Aerospace Engineering, Hong Kong University of Science and Technology, Kowloon, Hong Kong SAR}

\author{Ruhao Huang}
\affiliation{Department of Mechanical and Aerospace Engineering, Hong Kong University of Science and Technology, Kowloon, Hong Kong SAR}

\author{Xian Chen}
\email{xianchen@ust.hk}
\affiliation{Department of Mechanical and Aerospace Engineering, Hong Kong University of Science and Technology, Kowloon, Hong Kong SAR}


\date{\today}

\begin{abstract}
The ferroelectric material usually exhibits temperature dependent spontaneous polarization, known as pyroelectricity, which can be used to directly convert thermal energy to electricity from ambient low-grade waste heat. When utilizing the structural phase transformations of the material, the conversion capability can be magnified, consequently the device performance can be strongly boosted by orders of magnitude. However, common ferroelectric oxides suffer the mechanical fatigue and functional degradation over cyclic phase transformations, hindering widespread applications of the energy conversion device. In this paper, we investigate the mechanical and functional reversibility of the material by lattice tuning and grain coarsening. We discover the lead-free compound Ba(Ce$_{0.005}$Zr$_{0.005}$)Ti$_{0.99}$O$_{3}$-0.10(Ba$_{0.7}$Ca$_{0.3}$)TiO$_{3}$ (BCZT-0.10BCT) satisfying the compatibility condition among all present phases by its lattice parameters, making the phase transformations highly reversible. We demonstrated that the energy conversion device with the equiaxial coarse grains exhibits exceptional fatigue-resistance, with stable pyroelectric current output at 4$\mu$A/cm$^2$ over 3,000 energy conversion cycles. Our work opens a new way to fabricate high-performance material that advances the pyroelectric energy conversion for practical application in engineering.
\end{abstract}


\maketitle

\section{Introduction}
The ferroelectric materials that are easily polarized and show spontaneous polarization, are widely used for microelectronic sensors, transducers and super capacitors.  \cite{shin2017ferroelectric,gao2020review,kanno2018piezoelectric,butler2015ferroelectric}. Commonly, the spontaneous polarization of these polar materials is temperature dependent, known as the pyroelectric effect. Utilizing this effect, the ferroelectric materials can be used to convert energies from heat to electricity directly, which shows great potential in energy harvesting from low-grade waste heat ($< 150^\circ$C) \cite{zhang2019power,bucsek2019direct,zhang2021energy,zhang2021low}. Recent advances in pyroelectric energy conversion suggest that the phase transformation drives the abrupt change in polarization between the ferroelectric (FE) and paraelectric (PE) phases, thus strongly boosting the electricity generation from the small temperature fluctuations near the transition temperature  \cite{bowen2014pyroelectric,  bucsek2019direct,bucsek2020energy, zhang2021energy, zhang2021low}. The frequent thermal exchanges push the core material transforming reversibly between different crystal structures. This conversion is a ferroic, thermal and mechanical coupled process. Therefore, the loss of transformability and the accumulation of thermal hysteresis over successive energy conversion cycles become the critical issues that prevent it from wide applications and commercialization.
	
An effective method to enhance phase reversibility is to make the transforming material satisfy the compatibility conditions between phases \cite{song2013enhanced, chen2013study}. When the lattice parameters are tuned to satisfy the primary compatibility condition $\lambda_2 = 1$ where $\lambda_2$ is the middle eigenvalue of the transformation stretch tensor, the crystal structures of different symmetries can fit together through a stress-free interface during the phase transformation.\cite{zhang2009energy,ball1989fine,bhattacharya2003microstructure, chen2013study} The compatibility condition underlies a material development strategy to suppress functional degradation by lattice parameter tuning, which can be implemented by doping elements with similar physical properties but slightly different ionic radii. The method was initially introduced to develop martensitic alloys \cite{song2013enhanced,cui2006combinatorial,gu2018phase}, then widely accepted to discover the low hysteresis and low functional fatigue transforming ceramics and oxides \cite{pang2019reduced, jetter2019tuning, gu2021exploding}. Several studies have shown that the thermal hysteresis of functional oxides is minimized, simultaneously corresponding to enhanced transport properties when the $\lambda_2$ value of the material is approaching to 1. \cite{wegner2020correlation, liang2020tuning,zhang2021low, zhang2021energy} 
	
Theoretically, achieving the kinematic compatibility by lattice design is closely related to the single crystal behavior under the free boundary conditions. While the material is fabricated into a device integrated with electrodes under the application of thermal, mechanical and electric coupled loading, the overall performance of the device may depend on the microstructure, crystalline defects and orientations of the material. In most cases, the device exploits the polycrystalline material, whose  grain boundary may influence the multiferroic responses. Some experimental observations suggest that the coarse grains facilitate the domain switching \cite{cao1996grain,li2016effect}. Over thermally driven transformation cycles, the ferroelectrics with coarse grains usually exhibit a higher phase stability, relatively less electric leakage and a better pyroelectric property \cite{zhang2005computational,liu2015losses, zhang2020impact, zhang2019enhanced,yingwei2020influence, mao2014effect,chen2020grain, moya2013giant,huan2014grain,hanani2019enhancement}. Besides the ferroic properties, it is important to understand and explore how the grain morphology influences the structural, thermal, and mechanical reversibility under stress-induced transformation. In this paper, we investigate the lead-free ferroelectrics for pyroelectric energy conversion from two perspectives: 1. lattice compatibility, 2. grain morphology. We choose the compound system of barium titanate and barium calcium titanate as the starting point for lattice compatibility tuning. First, we aim at searching an optimal composition such that the phase transformations are highly reversible. Then, we study the effect of microstructure of grains on mechanical reversibility and functional reversibility for energy conversion device. Through the studies, we hope to advance the material development strategy of phase-transforming lead-free ferroelectric materials for energy conversion device fabrication and application.
 
\section{Lead-free ferroelectric material development}

The phase-transforming ferroelectric system Ba(Ce$_{0.005}$Zr$_{0.005}$)Ti$_{0.99}$O$_{3}$--$x$Ba$_{0.7}$Ca$_{0.3}$TiO$_{3}$ (abbreviated as BCZT-$x$BCT) is tuned by varying the atomic ratio $x \in [0.08, 0.12]$ at\%. The doping elements Ce and Zr were used to suppress the electric leakage \cite{zhang2020impact} and to improve the pyroelectric response \cite{zhang2019power, wegner2020correlation}.  
The ferroelectric oxides BCZT and BCT are synthesized by solid-state reaction \cite{zhang2020impact, zhang2021energy} respectively. See the Methods of Supplemental Materials for details.
The atomic composition of BCT is determined to be 0.10 at\%, i.e. BCZT-0.10BCT, whose lattice parameters closely satisfy the primary condition of compatibility for the phase transformation between ferroelectric and paraelectric phases. Among its nearby compositions between 0.08BCT and 0.12BCT, it shows the lowest thermal hysteresis \cite{zhang2021energy,zhang2020impact}, as seen in Figure S1 of Supplemental Materials.

To achieve different grain morphology, we utilized various sintering methods to sinter the two identical rod-shaped green bodies of BCZT-0.10BCT (See Methods of Supplemental Materials). One was conventionally sintered using a tube furnace \cite{dupuy2019effect}, while the other was directionally sintered using a 4-mirror infrared image furnace. The green body was sintered through a controlled grain growth by migrating the 4-mirror confocal plane slowly. Consequently, a gradual densification was achieved along a uniaxial direction. At a linear migration speed of 3mm/hour, the grains grow much coarser compared to those sintered using the high temperature tube furnace. After sintering, both rods were transversely cut into thin slices with a thickness of about 1 mm, which were well polished and chemically etched by 37\% hydrochloric acid to reveal the grain boundaries. The grain morphology of the sintered rods is studied by the polarized light-reflected differential interference microscope (DInM). For a polar material, the reflectivity depends on the grain orientation, by which we can develop a morphology map of grains. We analyzed the chemical composition and the crystal structure of the specimens by EDS and XRD experiments. The lattice parameters of cubic and tetragonal phases as well as the compositions are sufficiently identical between these two specimens sintered by different methods.  
	
\section{Results and discussion}

\subsection{Correlation between grain morphology and transport properties}

\begin{figure*}
\centering
\includegraphics[width = 0.65\textwidth]{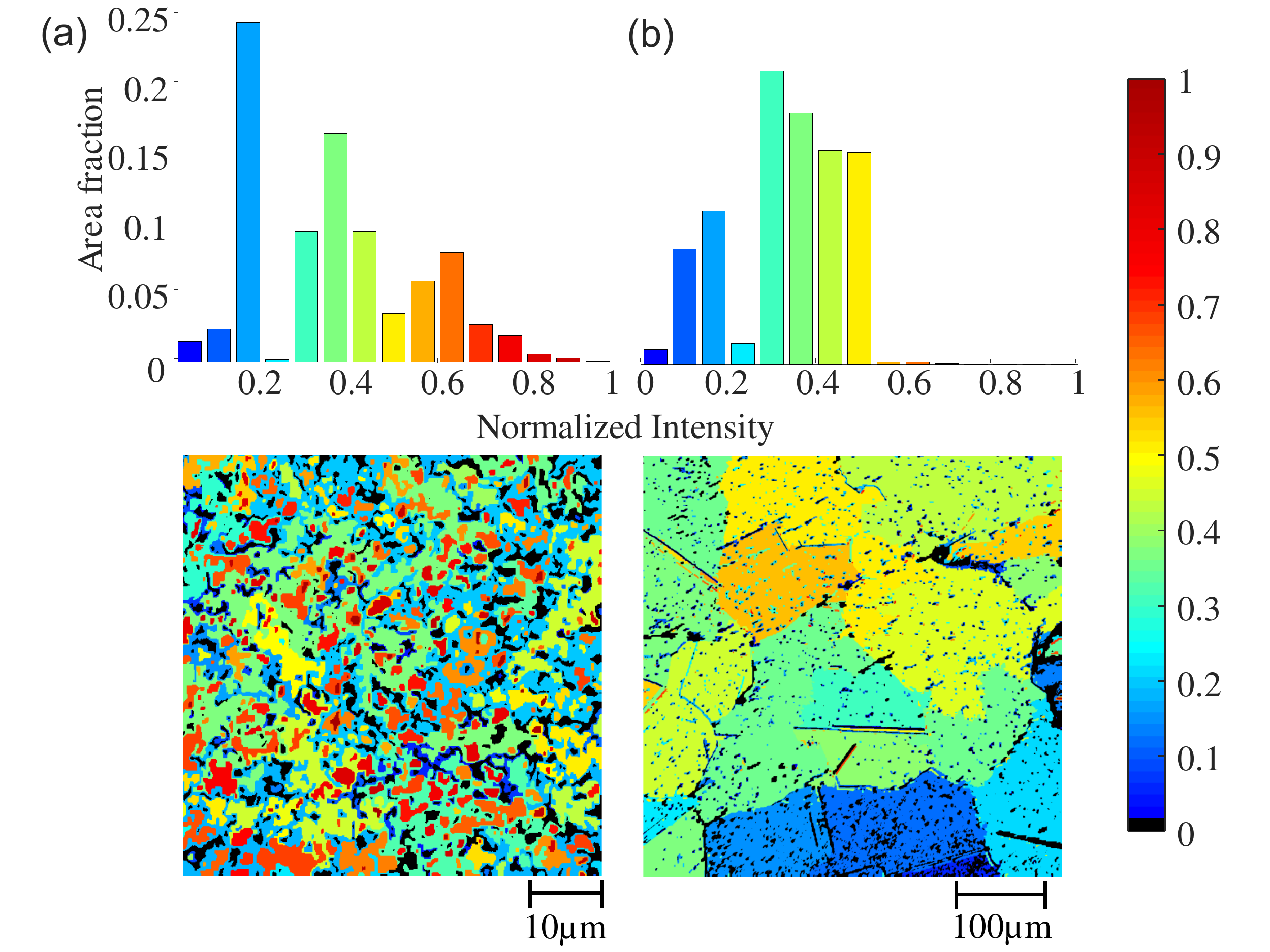}
\caption{The histogram of grain morphology and corresponding topographic maps of the BCZT-0.10BCT ferroelectrics sintered by (a) the conventional and (b) directional sintering methods respectively. }\label{Fig_grain}
\end{figure*}

Figure \ref{Fig_grain} shows the grain morphology of BCZT-0.1BCT observed by DInM, analyzed by the quantitative topographic mapping \cite{zeng2020dual}, for specimens with fine grains (FG) and coarse grains (CG). The colormap of micrographs indicates grains with varying orientations. We studied the statistics of the areal distribution of the grains, shown as the histograms in Figure \ref{Fig_grain}. While the specimen with fine grains exhibits a wide range of orientations with a random distribution, the coarse grain specimen has a preferred orientation for most of its grains. The grain size was presented as the mean diameter of each of the grains sampled from the micrographs. The average grain size is calculated as 1.7 $\mu$m for the FG sample and 160 $\mu$m for the CG sample. This result confirms the effectiveness of the directional sintering method to achieve the coarse grain ferroelectric material.

\begin{figure}
\centering
\includegraphics[width = 0.45\textwidth]{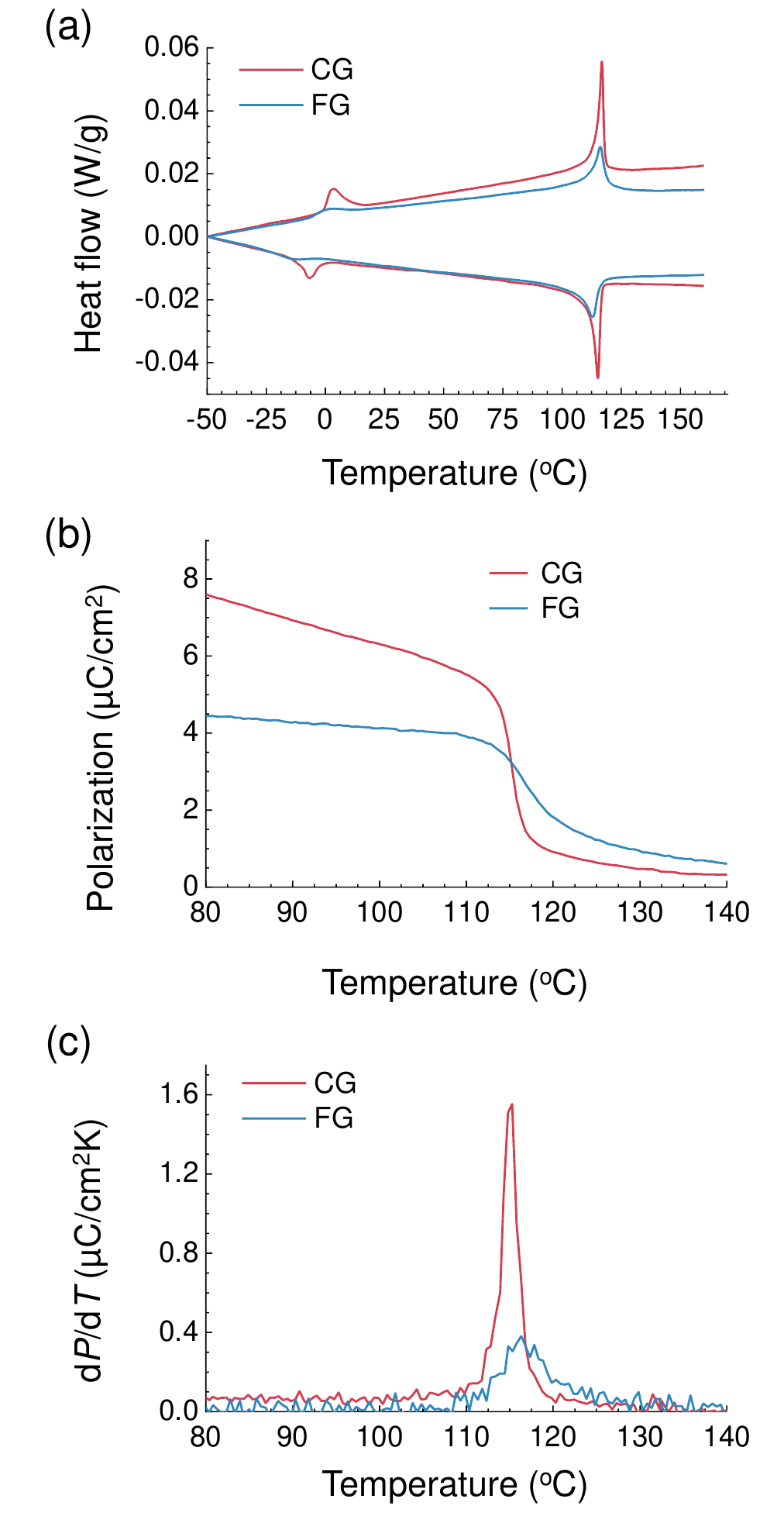}
\caption{The comparison of (a) thermal properties, (b) polarization and (c) pyroelectric coefficient versus temperature between the fine grains and the coarse grains BCZT-0.10BCT samples.}\label{Fig_PT}
\end{figure}

Figure \ref{Fig_PT} shows that the thermal and transport properties vary drastically as the grain morphology changes. Particularly, the phase transformation features are enhanced in the material with coarse grains. In the CG sample, we can clearly identify two reversible martensitic transformations in Figure \ref{Fig_PT} (a), that is the orthorhombic to tetragonal transformation at 0$^\circ$C and the tetragonal to cubic transformation at 115$^\circ$C. The tetragonal to cubic transformation is a first-order phase transformation corresponding to the ferroelectric to paraelectric transition as seen in Figure \ref{Fig_PT} (b) - (c). The latent heat of tetragonal to cubic transformation is measured as 0.665 J/g for the coarse grain sample and 0.514 J/g for the fine grain sample. The heat exchange due to the phase transformation is about 30\% more efficient in the sample with coarse grains, but the transition temperatures are the same for both CG and FG samples. This confirms that the lattice compatibility condition does not depend on the grain morphology, so does the thermal hysteresis, which is around 2$^\circ$C for both samples. The transition temperature $T_c = 115^\circ$C is usually denoted as the Curie temperature of the ferroelectric material. Near the Curie temperature, the transport properties such as the polarization ($P$) and pyroelectric coefficient ($dP/dT$) change abruptly, which also suggests that the material undergoes a first-order phase transformation. The spontaneous polarization in CG sample is almost twice as that in FG sample, and the pyroelectric coefficient of the CG sample is 1.55 $\mu$C/cm$^2$K, 
nearly enhanced by 4 times than that of the FG sample.
According to the performance analysis of the pyroelectric energy conversion device \cite{bucsek2019direct, bowen2014pyroelectric, zhang2019enhanced}, the CG sample is more suitable for the energy conversion by small temperature differences near the Curie temperature.  
	
\subsection{Mechanical and functional reversibility of the transforming ferroelectric materials}

\begin{figure*}
\includegraphics[width = 0.8\textwidth]{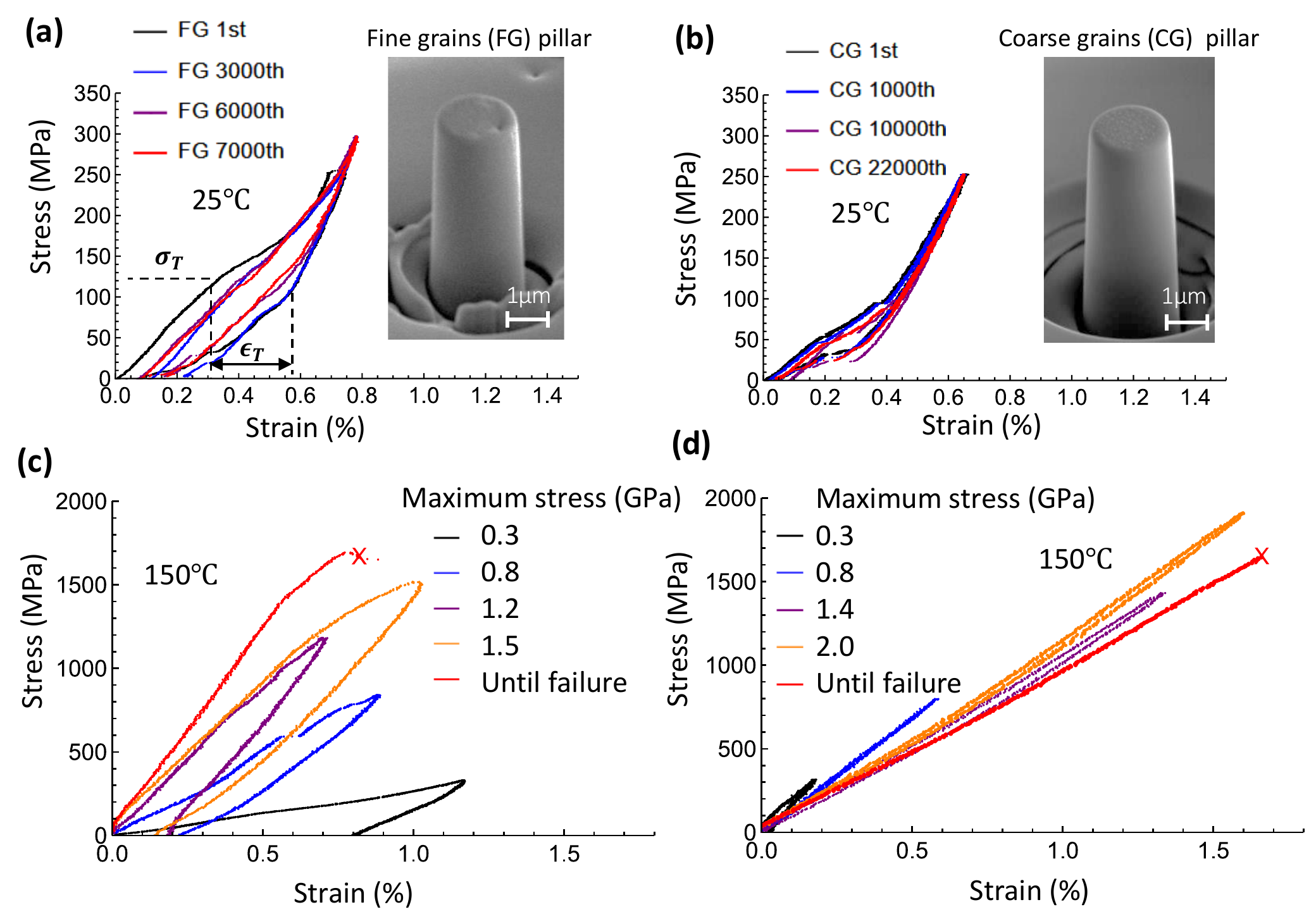}
\caption{The stress-strain curves of micropillars fabricated on (a) fine grains (FG) and (b) coarse grains (CG) at 25 $^\circ$C in ferroelectric phase over uniaxial microcompression cycles. The insets are the SEM images of the undeformed micropillars. The stress-strain curves of (c) the FG micropillar and (d) the CG micropillar under uniaxial compression at 150 $^\circ$C in paraelectric phase as increasing the maximum compressive stress until structural failure.}\label{Fig_fib}
\end{figure*}

The mechanical reversibility and durability of the transforming oxides were studied by the micropillar compression experiments for the FG and CG samples respectively. The cylindrical pillars of 2$\mu$m diameter were fabricated by focused-ion beam (FIB) milling \cite{karami2020two}, as shown in Figure \ref{Fig_fib}(a) for final grains and (b) for coarse grains. Note that the grain size of the CG sample is much larger than the pillar diameter, the micropillar in Figure \ref{Fig_fib}(b) is a single crystal, while the average grain size of the FG sample is slightly smaller than $2 \mu$m. The FG micropillar may consist of several grains. The microcompression cycles were carried out at 0.5Hz by the force control at room temperature by Hysitron TI 980 TriboIndenter. In each of the loading/unloading cycles, the pillar was loaded up to 800 $\mu$N, then completely unloaded. The experimental details are included in Methods of Supplemental Materials.

As seen in Figure \ref{Fig_fib}(a) - (b) for the stress-strain curves of the micropillars under cyclic loading, the micropillars of both fine grains and coarse grains are mechanically reversible over thousands of loading cycles, providing 0.7\% recoverable strains under approximately 300 MPa compressive stress. The mechanical response of the ferroelectric micropillars is nonlinear, which exhibit superelastic behaviors analogous to most of metallic martensitic materials \cite{manjeri2010, karami2021nanomechanics}, transforming intermetallic compounds \cite{sypek2017} and oxides \cite{lai2013shape, li2021, camposilvan2016size}. The superelastic strain can be obtained from the nonlinear portion of the stress-strain response as the strain $\epsilon_T$ in Figure \ref{Fig_fib} (a) corresponding to the driving stress $\sigma_T$. For the FG micropillar, $\sigma_T = 120$ MPa, $\epsilon_T= 0.27\%$. For the CG micropillar, the driving stress, $\sigma_T = 50$MPa is much lower, associated with slightly reduced superelastic strain $\epsilon_T = 0.2\%$. For tetragonal to orthorhombic transformation, the transformation stretch tensor can be calculated as $\sqrt{\mathbf C} \in \mathbb R^{3 \times 3}$ where $\mathbf C$ denotes the Right-Hand Cauchy Green tensor that deforms the tetragonal lattice to orthorhombic lattice. Depending on the symmetry relation and lattice parameters, there are two possible transformation mechanisms based on the weak Cauchy-Born Rule \cite{bhattacharya2003microstructure}, calculated as

{\small \begin{equation}\label{eq:C}
    \mathbf C = \left\{\renewcommand{\arraystretch}{0.8}\begin{array}{l}\begin{pmatrix}\frac{b_o^2+c_o^2}{4a_t^2} & 0 & \frac{-b_o^2 + c_o^2}{4 a_t c_t}\\0& \frac{a_o^2}{a_t^2}&0\\\frac{-b_o^2 + c_o^2}{4 a_t c_t} &0 & \frac{b_o^2+c_o^2}{4c_t^2}\end{pmatrix}, \text{ if mechanism I,}  \\\begin{pmatrix}\frac{b_o^2+c_o^2}{4a_t^2} & \frac{-b_o^2 + c_o^2}{4 a_t^2} & 0\\\frac{-b_o^2 + c_o^2}{4 a_t^2} & \frac{b_o^2+c_o^2}{4a_t^2} & 0 \\ 0 & 0 & \frac{a_o^2}{c_t^2}\end{pmatrix}, \text{ if 
mechanism II.}\end{array}\right.
\end{equation}}
Here, $a_t = 3.997\AA$ and $c_t = 4.0314\AA$ are the lattice parameters of the tetragonal phase measured by XRD (see Supplemental Materials) and $a_o = 5.675\AA$, $b_o = 5.69\AA$ and $c_o = 3.987\AA$ are the lattice parameters of the orthorhombic phase \cite{rhodes1951barium}. For mechanism I, the tetragonal a-axis and the two lateral face diagonals transform to the primitive unit cell of the orthorhombic lattice. The lattice correspondence is 
\begin{equation}\label{eq:m1}
   [101]_t || [100]_o, \ [010]_t || [100]_o, \ [\bar101]_t || [001]_o .
\end{equation}
For mechanism II, the tetragonal 4-fold c-axis and the basal face diagonals transform to the primitive unit cell of the orthorhombic lattice. The lattice correspondence is
\begin{equation}
     [110]_t || [001]_o,\ [1\bar10]_t || [010]_o, \ [001]_t || [100]_o. 
\end{equation} 

The mean transformation strain is calculated as 
{\small \begin{equation}
\epsilon_{\rm to} = \frac{1}{3}\sqrt{(\sqrt{\lambda_1}-1)^2 + (\sqrt{\lambda_2} -1)^2 + (\sqrt{\lambda_3}-1)^2}
\end{equation}}
where $\lambda_{1, 2, 3}$ are the eigenvalues of $\mathbf C$ given in \eqref{eq:C}. Direct calculation by substituting the lattice parameters gives $\epsilon_{\rm to} = 0.00231$ for mechanism I, and $\epsilon_{\rm to} = 0.00446$ for mechanism II. Compared to the measured superelastic strain from FG polycrystalline pillar (i.e. $0.27\% = 0.0027$), it highly suggests that the superelastic response of the ferroelectric micropillar is attributed to the reversible martensitic transformation by mechanism I in \eqref{eq:m1}. 

We calculated the middle eigenvalue of the transformation mechanism I as $\sqrt{\lambda_2} = 0.9977 \sim 1$, which satisfies the compatibility condition closely.
It suggests that the tetragonal to orthorhombic transformation of this material is as compatible as the cubic to tetragonal transformation at the Curie temperature. For the single crystal pillar in CG sample, it reversibly exhibits the superelasticity over 22,000 compression cycles without any nominal degradation, shown in Figure \ref{Fig_fib}(b). In contrast, the functional degradation is observed in the polycrystalline pillar with the same level of crystallographic compatibility, shown as the loss of reversibility and the decrease of superelastic strain upon 7000 mechanical cycles, in Figure \ref{Fig_fib}(a). 

Above the Curie temperature, we performed the micromechanical tests to study the overall superelasticity achieved by cubic to tetragonal and tetragonal to orthorhombic transformations. We adopted the same FIB milling process to fabricate micropillars for both FG and CG samples respectively. The uniaxial microcompression tests were conducted at 150$^\circ$C under the force control by the Triboindenter (Hysitron TI 980). Each of the micropillars was loaded to a preset maximum stress, then completely unloaded. Figure \ref{Fig_fib}(c) - (d) show the stress-strain behaviors of the micropillars as the preset maximum stress sequentially increases from 0.3GPa until failure. The mechanical properties of paraelectric FG pillar and CG pillar are drastically different. The FG micropillar exhibits a large stress hysteresis and irreversible strain upon each of the loading cycles. Finally it fails at 1.6GPa stress with 0.8\% ultimate strain. In contrast, the CG micropillar with the same composition and crystal structure shows an elastic behavior under all loading cycles, finally fails at 1.9GPa. It is striking that the CG pillar shows 1.6\% recoverable strain under nearly 2GPa compressive stress, which is almost twice as the ultimate deformation achieved by the FG pillar. Considering the entire structural transformations from cubic, to tetragonal and to orthorhombic phases, our experiment discovers that the grain morphology plays a significant role in mechanical properties and reversibility of the lead-free BCZT-BCT system. 

\subsection{Functional reversibility of energy conversion device}

\begin{figure*}
 \centering
\includegraphics[width = 0.9\textwidth]{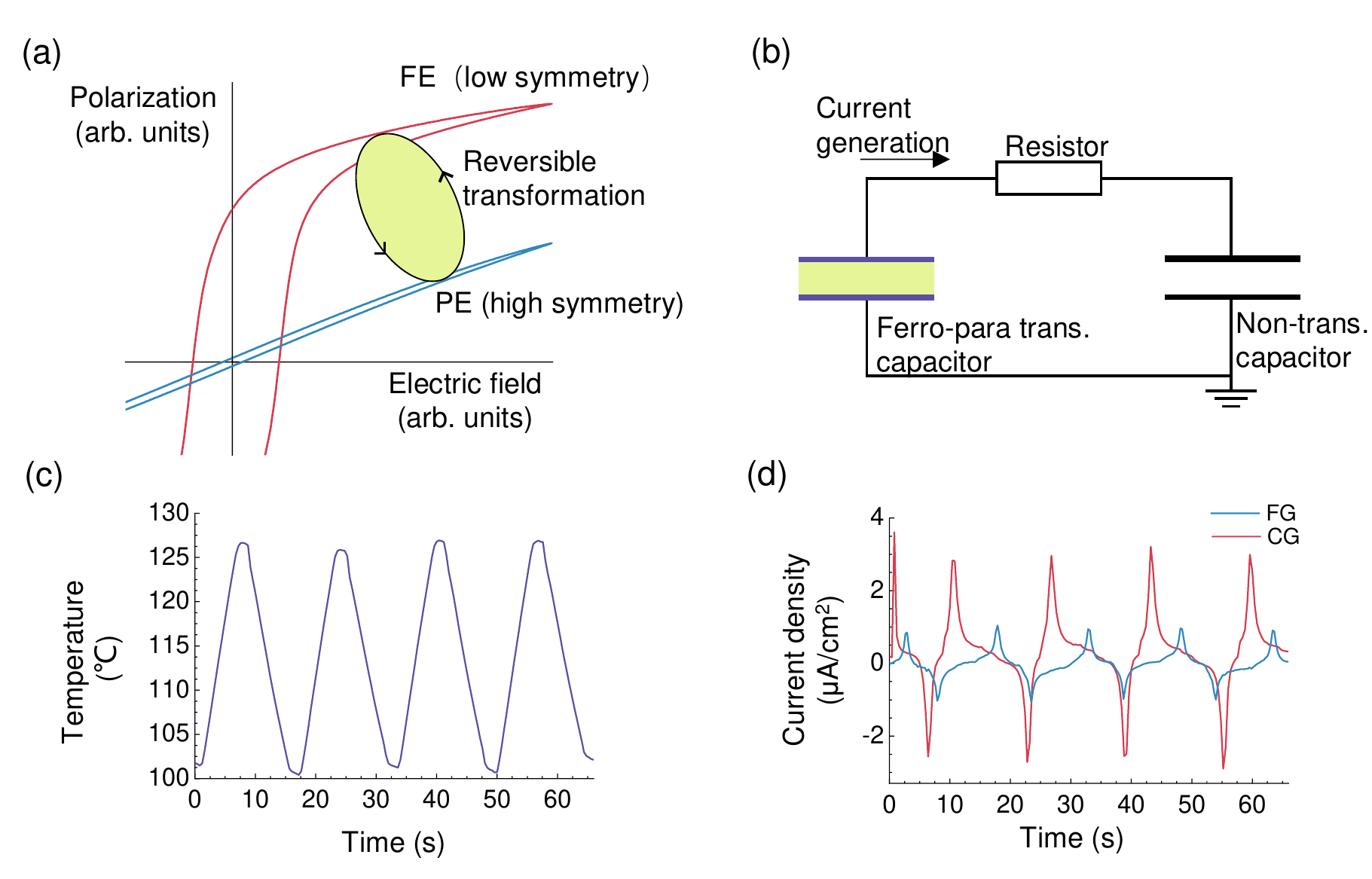}
\caption{Demonstration of bias-free pyroelectric energy conversion in BCZT-0.10BCT material. (a) Schematics of the bias-free thermodynamic model \cite{zhang2019power} with (b) the energy pick-up circuit to collect generated electricity. (c) Temperature fluctuations between 100$^\circ$C and 125$^\circ$C that fully cover the phase transformation between ferroelectric and paraelectric phases. (d) The measured electricity generated by FG and CG samples in the four consecutive conversion cycles.}\label{Fig_demo}
\end{figure*}

The influence of grain morphology on reversibility over the thermally-driven energy conversion cycles was studied by the pyroelectric energy conversion device \cite{zhang2021energy}. The conversion device is particularly designed to harvest the small temperature fluctuations near the Curie temperature of the first-order phase transformation between cubic paraelectric (PE) and tetragonal ferroelectric (FE) phases.  Figure \ref{Fig_demo} (a) elaborates the schematics of thermodynamic cycle based on the giant pyroelectric effect in the vicinity of the first-order phase transformation \cite{zhang2019power, bucsek2019direct}. This energy conversion system is different from conventional pyroelectric devices \cite{bowen2014pyroelectric}, as the thermodynamic system does not require an external voltage source to bias the isobaric process at different electric fields for the FE to PE path and the PE to FE path. The energy conversion device was initialized at a polarized state. The detailed device setup is included in the Supplemental Materials. The energy pick-up circuit is given in Figure \ref{Fig_demo} (b)\cite{zhang2021energy}, with the input thermal energy profile shown in Figure \ref{Fig_demo}(c) that fluctuates between 100$^\circ$C and 125$^\circ$C. The transition temperature of both samples was characterized as 115$^\circ$C (Figure \ref{Fig_PT}), which is fully covered by the temperature fluctuations.

The material performance of our energy conversion is evaluated by the Figure-of-Merit \cite{zhang2019power,zhang2021energy}
\begin{equation}\label{eq:fom}
{\rm FOM} = \frac{\llbracket P \rrbracket \kappa}{\ell},
\end{equation} 
in which $\kappa$ = $|dP/dT|_{max}$ is the steepest slope of the polarization with respect to temperature, $\llbracket P \rrbracket$ is the change of
polarization before and after phase transformation, and $\ell$ is the latent heat of phase transformation. The FOM of BCZT-0.10BCT was calculated from the thermal and ferroelectric characterization given by Figure \ref{Fig_PT}, that is 2.44$\mu$C$^2/$JcmK for the coarse grains and 0.45$\mu$C$^2/$JcmK for the fine grains. Consequently, the pyroelectric current density in Figure \ref{Fig_demo} (d) is measured as $4.0\mu$A/cm$^2$ for the coarse grains and $0.8\mu$A/cm$^2$ for the fine grains, which confirms that the grain coarsening enhances the pyroelectric energy conversion performance for the same material system by a factor of 5, which is consistent with the FOM evaluation in \eqref{eq:fom}.

 \begin{figure*}
\includegraphics[width = 0.9\textwidth]{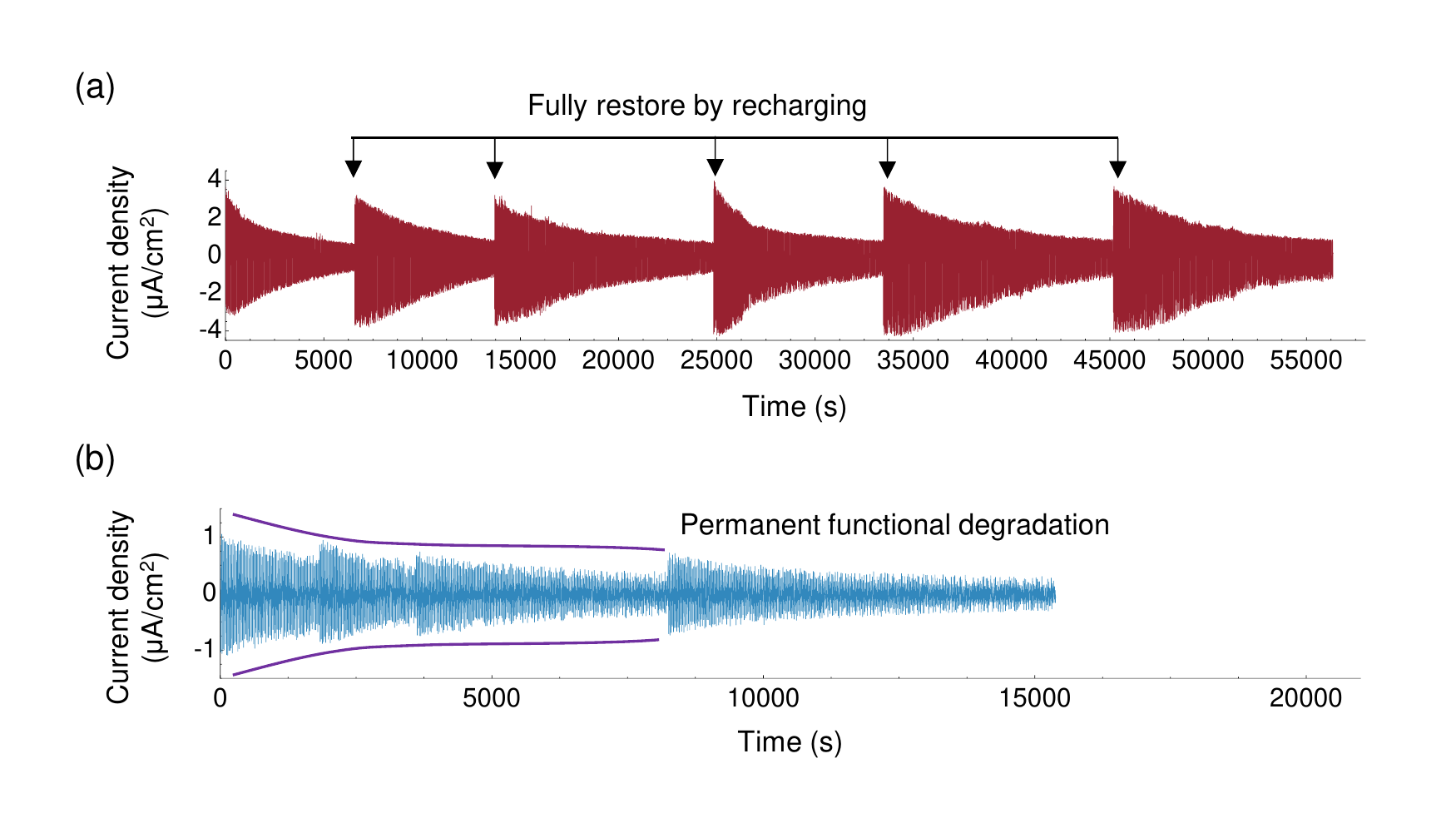}
\caption{ Characterization of reversibility over thermal cycles by the bias-free pyroelectric energy conversion in BCZT-0.10BCT material of different grain sizes. (a) 3,392 cycles of energy conversion in coarse grain (CG) material showing no functional degradation over 55,000s; (b) functional degradation in the fine grain (FG) material shows decreased performance.}\label{Fig_longdemo}
\end{figure*}

The functional reversibility over long energy conversion cycles is characterized in Figure \ref{Fig_longdemo}. As the developed ferroelectric compound satisfies the lattice compatibility for both ferroelectric and paraelectric phases, both FG and CG pyroelectric devices exhibit good functional reversibility upon thousands of conversion cycles. Specifically, the coarse grain sample with better mechanical reversibility and lower hysteresis stably and continuously generate approximately $4\mu$A/cm$^2$ pyroelectric current density over 3,000 conversion cycles lasting about 15 hours, in Figure \ref{Fig_longdemo}(a). As seen, the converted electric energy decades over time due to the unavoidable electric leakage of the dielectric material \cite{zhang2020impact}, but this is not the intrinsic fatigue of the functional material. During conversion, we just need to reset the device periodically by recharging the transforming pyroelectric capacitor to its initial state, called an e-pack period \cite{zhang2021energy}. As a result, the energy generation capacity is completely reversed to the reference value if the phase-transforming material sustains its functionality. Upon five e-pack periods, Figure \ref{Fig_longdemo}(a) shows that the coarse grain pyroelectric device still generates the pyroelectric current at the same level as in its first few cycles. However, the fine-grain pyroelectric device shows a clear trend of functional degradation besides the electric leakage in Figure \ref{Fig_longdemo}(b). That is the reference pyroelectric current becomes smaller and smaller in each of the e-pack periods. Finally, the functionality degrades to half of the pyroelectric current produced in the first cycle after 15,000 seconds (i.e. 4 hours), which implies that the phase-transforming ferroelectric layer of the pyroelectric device has irreversibly lost its energy conversion capability. Table \ref{tab_comparison} summarizes the mechanical and functional properties of both FG and CG pyroelectric energy conversion devices, suggesting that the crystallographic compatible CG material with approximately equiaxial 160$\mu$m sized grains is an ideal material candidate for pyroelectric energy conversion by phase transformation. 

 \begin{table}[h!]
 \caption{\label{tab_comparison} Comparison of the microstructure, mechanical and energy conversion properties of the BCZT-0.10BCT ferroelectric material.}
 \begin{ruledtabular}
 \begin{tabular}{lll}
\multirow{2}{*}{Properties} & \multicolumn{2}{c}{BCZT-0.10BCT}\\
& FG & CG \\\hline
Grain size ($\mu$m) &  1.67  & 160 \\
$\sigma_{\text{max}}$(GPa)  &1.70  & 1.91 \\
$dP/dT$ at $T_c$ ($\mu$C/cm$^2$K) &  0.38 & 1.55 \\
FOM ($\mu$C$^2/$J cm K) &  0.45 &2.44\\
$i_p$ ($\mu$A/cm$^2$) & 0.8 & 4.0\\
 \end{tabular}
 \end{ruledtabular}
 \end{table}

\section{Conclusion}

Energy harvesting from heat energies in the low-temperature regime (i.e. $< 200^\circ$C) is an interdisciplinary and multifaceted grand challenge in energy science. Rational design of functional materials is undoubtedly one of the key imperatives. In this paper, we introduced and demonstrated an effective development strategy to enhance the material properties for energy conversion at two levels: 1. to make lattice parameters of phases satisfy the kinematic compatibility condition by compositional tuning; 2. to coarsen the grains (i.e. over 150$\mu$m) effectively by directional sintering method. We proposed the lead-free BCZT-0.10BCT exhibits exceptionally good mechanical reversibility and high mechanical strength by satisfying the compatibility condition closely for the entire transformation regime. The pyroelectric device comprised of equiaxial coarse grains significantly enhances the energy conversion performance by nearly five times, compared to the fine grains device.
In addition, the coarse grain device are running over tens of thousands of transformation cycles with stable and continuous pyroelectric current output at the microAmpere level.  
Our work opens a way to effectively discover high-performance pyroelectric material with strong mechanical reversibility, which significantly advances the application and commercialization of the pyroelectric energy conversion by small temperature fluctuations.
	
 \begin{acknowledgments}
C. Z. acknowledge the support by National Natural Science Foundation of China (No. 12204350), Fundamental Research Funds for the Central Universities (No.22120220252) and Shanghai Leading Talents (Overseas) Program. X. C., Z. Z., K. H. C. and R. H. thank the financial support under GRF Grants 16203021, 16204022 and CRF Grant No. C6016-20G-C by Research Grants Council, Hong Kong. X.C. and C.Z. thank the support by Innovation and Technology Commission, HK through the  ITF Seed grant ITS/354/21.
\end{acknowledgments}

\bibliography{grainsizereversibility}

\end{document}